\documentclass[twocolumn,showpacs,prl,superscriptaddress]{revtex4-1}

\usepackage{graphicx}
\usepackage{epstopdf}
\usepackage{dcolumn}
\usepackage{bm}
\usepackage{color}
\usepackage{cancel}
\usepackage{ulem}

\begin{document}

\title{Azimuthal anisotropy in U+U and Au+Au collisions at RHIC}

\affiliation{AGH University of Science and Technology, Cracow 30-059, Poland}
\affiliation{Argonne National Laboratory, Argonne, Illinois 60439, USA}
\affiliation{Brookhaven National Laboratory, Upton, New York 11973, USA}
\affiliation{University of California, Berkeley, California 94720, USA}
\affiliation{University of California, Davis, California 95616, USA}
\affiliation{University of California, Los Angeles, California 90095, USA}
\affiliation{Central China Normal University (HZNU), Wuhan 430079, China}
\affiliation{University of Illinois at Chicago, Chicago, Illinois 60607, USA}
\affiliation{Creighton University, Omaha, Nebraska 68178, USA}
\affiliation{Czech Technical University in Prague, FNSPE, Prague, 115 19, Czech Republic}
\affiliation{Nuclear Physics Institute AS CR, 250 68 \v{R}e\v{z}/Prague, Czech Republic}
\affiliation{Frankfurt Institute for Advanced Studies FIAS, Frankfurt 60438, Germany}
\affiliation{Institute of Physics, Bhubaneswar 751005, India}
\affiliation{Indian Institute of Technology, Mumbai 400076, India}
\affiliation{Indiana University, Bloomington, Indiana 47408, USA}
\affiliation{Alikhanov Institute for Theoretical and Experimental Physics, Moscow 117218, Russia}
\affiliation{University of Jammu, Jammu 180001, India}
\affiliation{Joint Institute for Nuclear Research, Dubna, 141 980, Russia}
\affiliation{Kent State University, Kent, Ohio 44242, USA}
\affiliation{University of Kentucky, Lexington, Kentucky, 40506-0055, USA}
\affiliation{Korea Institute of Science and Technology Information, Daejeon 305-701, Korea}
\affiliation{Institute of Modern Physics, Lanzhou 730000, China}
\affiliation{Lawrence Berkeley National Laboratory, Berkeley, California 94720, USA}
\affiliation{Max-Planck-Institut fur Physik, Munich 80805, Germany}
\affiliation{Michigan State University, East Lansing, Michigan 48824, USA}
\affiliation{Moscow Engineering Physics Institute, Moscow 115409, Russia}
\affiliation{National Institute of Science Education and Research, Bhubaneswar 751005, India}
\affiliation{Ohio State University, Columbus, Ohio 43210, USA}
\affiliation{Institute of Nuclear Physics PAN, Cracow 31-342, Poland}
\affiliation{Panjab University, Chandigarh 160014, India}
\affiliation{Pennsylvania State University, University Park, Pennsylvania 16802, USA}
\affiliation{Institute of High Energy Physics, Protvino 142281, Russia}
\affiliation{Purdue University, West Lafayette, Indiana 47907, USA}
\affiliation{Pusan National University, Pusan 609735, Republic of Korea}
\affiliation{University of Rajasthan, Jaipur 302004, India}
\affiliation{Rice University, Houston, Texas 77251, USA}
\affiliation{University of Science and Technology of China, Hefei 230026, China}
\affiliation{Shandong University, Jinan, Shandong 250100, China}
\affiliation{Shanghai Institute of Applied Physics, Shanghai 201800, China}
\affiliation{Temple University, Philadelphia, Pennsylvania 19122, USA}
\affiliation{Texas A\&M University, College Station, Texas 77843, USA}
\affiliation{University of Texas, Austin, Texas 78712, USA}
\affiliation{University of Houston, Houston, Texas 77204, USA}
\affiliation{Tsinghua University, Beijing 100084, China}
\affiliation{United States Naval Academy, Annapolis, Maryland, 21402, USA}
\affiliation{Valparaiso University, Valparaiso, Indiana 46383, USA}
\affiliation{Variable Energy Cyclotron Centre, Kolkata 700064, India}
\affiliation{Warsaw University of Technology, Warsaw 00-661, Poland}
\affiliation{Wayne State University, Detroit, Michigan 48201, USA}
\affiliation{World Laboratory for Cosmology and Particle Physics (WLCAPP), Cairo 11571, Egypt}
\affiliation{Yale University, New Haven, Connecticut 06520, USA}
\affiliation{University of Zagreb, Zagreb, HR-10002, Croatia}

\author{L.~Adamczyk}\affiliation{AGH University of Science and Technology, Cracow 30-059, Poland}
\author{J.~K.~Adkins}\affiliation{University of Kentucky, Lexington, Kentucky, 40506-0055, USA}
\author{G.~Agakishiev}\affiliation{Joint Institute for Nuclear Research, Dubna, 141 980, Russia}
\author{M.~M.~Aggarwal}\affiliation{Panjab University, Chandigarh 160014, India}
\author{Z.~Ahammed}\affiliation{Variable Energy Cyclotron Centre, Kolkata 700064, India}
\author{I.~Alekseev}\affiliation{Alikhanov Institute for Theoretical and Experimental Physics, Moscow 117218, Russia}
\author{J.~Alford}\affiliation{Kent State University, Kent, Ohio 44242, USA}
\author{A.~Aparin}\affiliation{Joint Institute for Nuclear Research, Dubna, 141 980, Russia}
\author{D.~Arkhipkin}\affiliation{Brookhaven National Laboratory, Upton, New York 11973, USA}
\author{E.~C.~Aschenauer}\affiliation{Brookhaven National Laboratory, Upton, New York 11973, USA}
% ...

\collaboration{STAR Collaboration}\noaffiliation


\begin{abstract}
  Collisions between prolate uranium nuclei are used to study how
  particle production and azimuthal anisotropies depend on initial
  geometry in heavy-ion collisions. We report the two- and four-
  particle cumulants, $v_2\{2\}$ and $v_2\{4\}$, for charged hadrons
  from U+U collisions at $\sqrt{s_{\rm NN}}$ = 193 GeV and Au+Au
  collisions at $\sqrt{s_{\rm NN}}$ = 200 GeV. Nearly fully
  overlapping collisions are selected based on the energy deposited by
  spectators in Zero Degree Calorimeters (ZDCs). Within this sample,
  the observed dependence of $v_2\{2\}$ on multiplicity demonstrates
  that ZDC information combined with multiplicity can preferentially
  select different overlap configurations in U+U collisions. We also
  show that $v_2$ vs multiplicity can be better described by models,
  such as gluon saturation or quark participant models, that eliminate
  the dependence of the multiplicity on the number of binary
  nucleon-nucleon collisions.
\end{abstract}

\maketitle

Collisions of nuclei at the Relativistic Heavy-Ion Collider (RHIC) and
the Large Hadron Collider (LHC) create a fireball hot and dense enough
to form a Quark Gluon Plasma (QGP)~\cite{Gyulassy:2004zy}.
Anisotropies in the final momentum space distributions can be traced
back to spatial anisotropies in the initial state and are used to
understand the nature of the fireball
~\cite{Mishra:2007tw,Sorensen:2011hm}. These anisotropies are studied
using harmonics of the distribution of the azimuthal angle $\phi$
separation between pairs of
particles~\cite{Ollitrault:1992bk,Ackermann:2000tr,Adams:2005dq}.
The inference of the properties of the fireball from these
measurements is limited however by uncertainties in the
description of the initial state~\cite{Hirano:2005xf}. Collisions
between uranium nuclei, which have an intrinsic prolate
shape~\cite{Raman:1201zz}, provide a way to manipulate this initial
geometry to test our understanding of the initial state of heavy-ion
collisions and the subsequent fireball~\cite{UUpapers}.

Even in nearly fully overlapping collisions of U nuclei (impact
parameter $b\approx 0$ fm), the initial matter distribution can
exhibit very different shapes. In one extreme, the major axes of both
colliding nuclei could lie parallel to the beam so that the tip of one
nucleus impinges on the tip of the other (tip-tip). Another extreme
occurs if the major axes of the nuclei are parallel to each other but
perpendicular to the beam so that they collide side-on-side or
body-body. There are two principal differences in these two
configurations --- tip-tip collisions have a larger number of binary
nucleon-nucleon collisions $N_{\rm bin}$ while body-body collisions
have a smaller $N_{\rm bin}$ but a more elliptic overlap region
(larger eccentricity $\varepsilon_2$). The larger $N_{\rm bin}$ in the
tip-tip configuration is expected to lead to a larger multiplicity of
produced particles~\cite{Kharzeev:2000ph,Miller:2007ri} while the more
elliptic shape of the body-body collisions is expected to lead to a
larger second harmonic anisotropy $v_2$. The dependence of $v_2$ on
multiplicity in nearly fully overlapping U+U collisions therefore
tests our understanding of particle production and the development of
$v_2$. An anti-correlation between $v_2$ and multiplicity in these
collisions will also demonstrate that multiplicity can be used to
select enhanced samples of body-body or tip-tip configurations.  Those
samples can then be used to study other topics like the path-length
dependence of jet-quenching~\cite{UUpapers}, or the extent to which
three-particle charge-dependent
correlations~\cite{Abelev:2009ac,Sergei_PRL} can be attributed to
local parity violation~\cite{Kharzeev:2004ey} or background
effects~\cite{Schlichting:2010qia}. We also investigate two models
that do not include any explicit dependence on $N_{\rm bin}$; one
based on gluon saturation~\cite{McLerran:1993ni,Schenke:2014tga} and
the other based on the number of participating constituent
quarks~\cite{Eremin:2003qn,Adler:2013aqf}.

In this Letter we report measurements of the two- and four-particle
cumulant of $v_2$ ($v_2\{2\}$ and $v_2\{4\}$) in $^{197}$Au+$^{197}$Au
and $^{238}$U + $^{238}$U collisions at $\sqrt{s_{\rm NN}}= 200$ and
193 GeV respectively.  Both minimum bias and nearly fully overlapping
events where most of the nucleons participate in the collision are
studied.  The data sets were collected by STAR~\cite{Ackermann:2002ad}
in 2011 and 2012. The U+U data consists of approximately 307 million
events including 7 million specially triggered central events. Charged
particles within pseudo-rapidity window $|\eta|<1$ were detected using
the STAR Time Projection Chamber (TPC)~\cite{Anderson:2003ur}. We
select tracks within the transverse momentum range $0.2 < p_{T} < 2.0
$ GeV/$c$. The STAR Zero-degree Calorimeters
(ZDCs)~\cite{Bieser:2002ah} were used to select the sample of nearly
fully overlapping events; those having large multiplicity but little
activity in the ZDCs. The ZDC resolution was determined to be $23 \pm
2$\% from the observation of the single neutron peak in the ADC
signal. The ZDC selection requires ZDCs on both sides of the detector
to have a signal smaller than the specified cut.
The tracking efficiency is corrected via embedding and weights in
$\eta$ and $\phi$ derived from the inverse of the distribution of
tracks observed over many events. This method allows us to correct our
$v_2\{2,4\}$ measurements for imperfections in the tracking
efficiency. $v_2\{4\}$ was calculated using the Q-Cumulant
method~\cite{Bilandzic:2010jr} while $v_2\{2\}$ was calculated
directly from particle pairs $\langle\cos2(\phi_1-\phi_2)\rangle$. To
reduce the contribution from HBT, Coulomb and track-merging effects, a
minimum $\eta$ separation of $|\Delta\eta| > 0.1$ is required for
$v_2\{2\}$. Measurement uncertainties were estimated by varying event
and track selection criteria, varying efficiency estimates, and by
comparing data from different run periods. These uncertainties are
quite small; less than 0.1\% absolute variation on $v_2\{2,4\}$.

\begin{figure}
\includegraphics[width=3.2in]{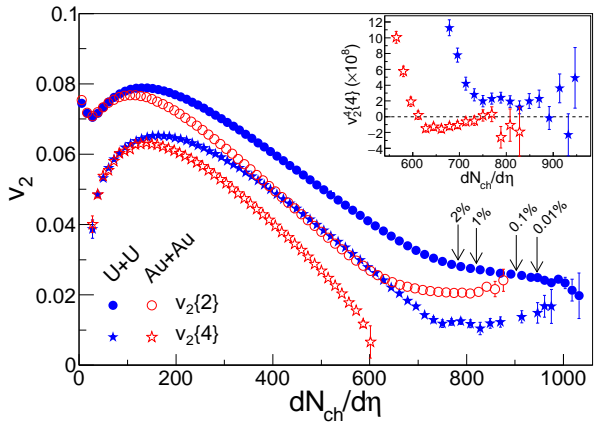}
\caption{\label{fig:fig01}(Color online) The two- and four- particle
  cumulant $v_2\{2\}$ and $v_2\{4\}$ within $|\eta|<1$ versus
  $dN_{\rm ch}/d\eta$ from 200 GeV Au+Au and 193 GeV U+U
  collisions. Dashed lines show U+U centralities based on
  $dN_{\rm ch}/d\eta$ measured in $|\eta|<0.5$. $v_2^4\{4\}$ (the
  experimentally observed quantity) is shown in the inset without
  taking the fourth root in the range where it is near zero or
  negative.}
\end{figure}

Figure~\ref{fig:fig01} shows the two- and four- particle cumulant
$v_2\{2\}$ and $v_2\{4\}$ from minimum bias 200 GeV Au+Au and 193 GeV
U+U collisions as a function of efficiency corrected charged particle
multiplicity $dN_{\rm ch}/d\eta$. We find that the relationship
between $dN_{\rm ch}/d\eta$ and centrality fraction can be
parameterized as $(dN_{\rm ch}/d\eta)^{1/4} = c_1 - c_2x +
c_3\exp(-c_4x^{c_5})$ with $c_1$=5.3473, $c_2$=4.298, $c_3$=0.2959,
$c_4$=18.21, and $c_5=0.4541$ for U+U and $c_1$=5.0670, $c_2$=3.923,
$c_3$=0.2310, $c_4=18.37$, and $c_5$=0.4842 for Au+Au. Multiplicity
trends for $v_2\{2\}$ and $v_2\{4\}$ in U+U collisions are mostly
similar to those observed in Au+Au collisions. A notable difference
however is seen in the $v_2\{4\}$ measurements in central U+U
collisions. Whereas $v_2^4\{4\}$ (shown in the inset) is negative for
central Au+Au collisions, it is positive for U+U collisions. Previous
studies showed that fluctuations in the number of participating
nucleons cause $v_2^4\{4\}$ in central Au+Au collisions to become
negative~\cite{Agakishiev:2011eq}. The observation of $v_2^4\{4\}>0$
in the most central U+U collisions indicates that the prolate shape of
uranium increases the anisotropy in the final momentum space
distributions of the observed particles.

Glauber-based models have typically used a two-component model
($(1-x_{\rm hard})N_{\rm part}/2+x_{\rm hard}N_{\rm bin}$) for the
multiplicity, where $N_{\rm part}$ is the number of struck nucleons,
$N_{\rm bin}$ is the number of binary nucleon-nucleon collisions, and
$x_{\rm hard}$ is a fractional contribution of $N_{\rm bin}$ to the
multiplicity~\cite{Kharzeev:2000ph,Miller:2007ri}. The multiplicity is
then assumed to fluctuate according to a convolution of negative
binomial distributions (NBD) with parameters $n$ and $k$ related to
the mean and width measured from $p$+$p$ collisions at the same energy
and in the same $|\eta|$ window~\cite{Ansorge:1988kn}. We will refer
to this model as ``Glauber-$x_{\rm hard}$.'' Since the number of hard
scatterings is known to scale with $N_{\rm bin}$, $x_{\rm hard}$ is
often thought of as reflecting the contribution of hard processes to
the multiplicity. It can also be thought of as a coherence parameter
with $x_{\rm hard}=1$ giving the maximum incoherence as multiplicity
entirely arises from independent binary nucleon-nucleon
collisions. The Glauber-$x_{\rm hard}$ model indicates that $v_2$ in
U+U collisions should begin to decrease markedly for events with
multiplicities in the top 1\%~\cite{Sergei_PRL} forming a knee
structure where tip-tip collisions with larger $N_{\rm bin}$ and
smaller eccentricity begin to dominate. Vertical dashed lines in the
figure indicate the 1\%, 0.1\% and 0.01\% highest multiplicity U+U
collisions. No knee structure is observed suggesting the
Glauber-$x_{\rm hard}$ model may not be the correct
description. Adding more multiplicity fluctuations causes the knee
structure to disappear~\cite{Rybczynski:2012av} but this will also
significantly increase the average $\varepsilon_2$ in central
collisions.

\begin{figure}
\includegraphics[width=3.2in]{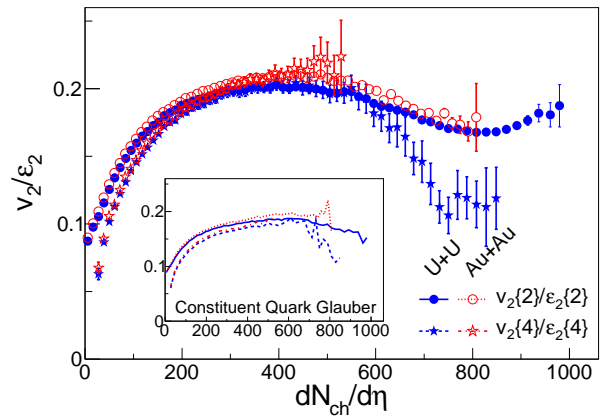}
\caption{\label{fig:fig02}(Color online) $v_2$ scaled by participant
  eccentricity from 200 GeV Au+Au and 193 GeV U+U collisions. The
  eccentricity distributions are calculated in a Monte Carlo Glauber
  model~\cite{Bhalerao:2006tp,Broniowski:2007ft,Masui:2009qk,NewUU}. Both
  U+U and Au+Au follow a similar trend for $v_2/\varepsilon_2$ and a
  turn-over is observed in central collisions.  The inset shows the
  same quantity but with the eccentricity calculated in a constituent
  quark Glauber model~\cite{Eremin:2003qn,Adler:2013aqf} with the
  Woods-Saxon parameters proposed in Ref.~\cite{NewUU}.}
\end{figure}

To explore the dependence of $v_2$ on the initial eccentricity
$\varepsilon_2$, we plot $v_2/\varepsilon_2$ versus $dN_{\rm
  ch}/d\eta$. It was found previously that $v_2/\varepsilon_2$
monotonically increases with increasing $dN_{\rm ch}/d\eta$ and
depending on the model for the initial eccentricity may, or may not
saturate in the most central
collisions~\cite{Agakishiev:2011eq}. Figure~\ref{fig:fig02} shows
$v_2\{2\}/\varepsilon_2\{2\}$ and $v_2\{4\}/\varepsilon_2\{4\}$ from
Au+Au and U+U collisions. $\varepsilon_2\{2\}$ and
$\varepsilon_2\{4\}$ are the second and fourth cumulants of the
participant eccentricity distributions calculated from the
Glauber-$x_{\rm hard}$
model~\cite{Bhalerao:2006tp,Broniowski:2007ft,Masui:2009qk}. Both U+U
and Au+Au follow a similar trend for $v_2/\varepsilon_2$. However, a
turn-over is observed in central collisions ($dN_{\rm ch}/d\eta>500$).
This has not been observed previously since measurements have
typically been integrated over 5\% most
central~\cite{Agakishiev:2011eq}. The turn-over is consistent with the
model overestimating $\varepsilon_2$ in central collisions.
Increasing the multiplicity fluctuations as in
Ref.~\cite{Rybczynski:2012av} will only increase the eccentricity in
central collisions suggesting that a different explanation may be
required to explain both the turn-over of $v_2/\varepsilon_2$ and the
lack of a knee structure in $v_2$ vs $dN_{\rm ch}/d\eta$. Using a new
set of Woods-Saxon parameters derived in Ref.~\cite{NewUU} with a
smaller diffuseness and smaller deformation parameter $\beta_2$ in
combination with the same Glauber model, reduces the downturn in
central U+U collisions somewhat but introduces a mismatch between the
U+U and Au+Au curves with the Au+Au curves higher while
$v_2\{4\}/\varepsilon_2\{4\}$ for U+U still exhibits a downturn (not
shown). In the inset of the figure, we show the result for a new
Glauber calculation using contituent quarks as
participants~\cite{Eremin:2003qn,Adler:2013aqf} and the new set of
parameters~\cite{NewUU}. This estimate for $\varepsilon_2$ leads to a
seemingly more natural behavior for $v_2/\varepsilon_2$ with the drop
in the highest multiplicity collisions almost entirely gone. The model
will be investigated and discussed further below.

\begin{figure}
\includegraphics[width=3.3in]{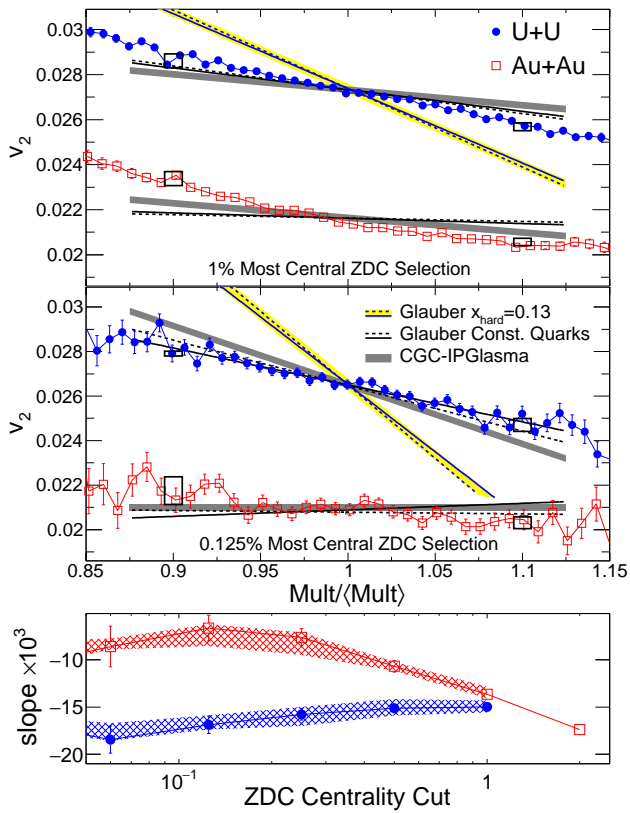}
\caption{\label{fig:fig03}(Color online) Top panels: charged particle
  $v_2\{2\}$ vs. normalized multiplicity within $|\eta| < 1.0$. The
  upper panel is for the top 1\% most central events based on the
  smallness of the ZDC signal, while the middle panel is for the top
  0.125\%.  Small boxes indicate the possible range of variation of
  $v_2$ from uncertainties in the efficiency corrections on the
  x-axis. Model comparisons are described in the text. Bottom panel:
  The slopes as a function of increasingly tighter ZDC centrality
  selections. The systematic uncertainties are shown as bands. }

\end{figure}

The trends of $v_2$ versus $dN_{\rm ch}/d\eta$ are mostly dominated by the
elliptic shape of the overlap region in collisions with a non-zero
impact parameter. To study body-body or tip-tip collisions we
investigate nearly fully overlapping collisions with minimal activity
in the ZDCs. If body-body collisions produce smaller multiplicities
than tip-tip collisions, we expect to see a negative slope in $v_2$ vs
multiplicity for these collisions. A negative slope, however, can also
come from contamination from larger impact parameter collisions. To
assess their contribution we use collisions of more spherical Au
nuclei as a control sample. Figure~\ref{fig:fig03} shows the elliptic
flow $v_2\{2\}$ of all charged particles as a function of the
normalized multiplicity (Mult/$\langle{\rm Mult}\rangle$) for two
different systems. We increase the acceptance to $|\eta| < 1.0$ to
reduce multiplicity fluctuations. The upper panel shows the results
for the 1\% most central events based on the smallest signal seen in
the ZDCs. Both Au+Au and U+U show a negative slope, which indicates
the effect of the impact parameter is still prominent (otherwise we
expect the Au+Au slope to be nearly flat or even positive). The middle
panel of Fig.~\ref{fig:fig03} shows the 0.125\% most central
events. The negative slope for Au+Au collisions is smaller in
magnitude, indicating the effects from non-central collisions are
reduced and the variation in multiplicity in Au+Au collisions is
mainly driven by fluctuations.  The bottom panel of
Fig.~\ref{fig:fig03} shows how the slopes extracted from $v_2$ vs
normalized multiplicity evolve with successively tighter ZDC
sections. While the slope for Au+Au collisions becomes less negative,
the slope for U+U collisions becomes steeper as the centrality
selection is tightened. This demonstrates that the variation of
multiplicity in the 0.125\% U+U collisions is dominated by the
different geometries made possible by the prolate shape of the uranium
nucleus and that tip-tip collisions produce more multiplicity than
body-body collisions.  Systematic uncertainties shown as bands on the
slope were estimated by varying the fit range and efficiency
corrections. Other sources of systematic error are smaller and
sub-dominant compared to the variation due to the range of
efficiencies used in the error analysis.  Due to large statistical
errors, no conclusions could be drawn from studies of $v_2\{4\}$
versus multiplicity in these events. We also measured $v_{3}\{2\}$ in
central collisions and found that $v_3\{2\}$ in the 0.125\% most
central collisions are $(1.410 \pm 0.006) \times 10^{-2}$ for U+U and
$(1.380 \pm 0.008)\times 10^{-2}$ in Au+Au collisions (statistical
errors only). The slope of $v_3$ vs multiplicity was small and
negative in both systems at about $-0.005\pm0.002$.

The U+U data in the top panels of Fig.~\ref{fig:fig03} are compared to
the Glauber-$x_{\rm hard}$ model (asssuming $v_2=\varepsilon_2\langle
v_2\rangle/\langle\varepsilon_2\rangle$). The ZDC response was modeled
by calculating the number of spectator neutrons from the Glauber model
(accounting for the charge to mass ratio of the nucleus) and folding
each neutron with the known ZDC resolution for a single neutron.
The Glauber-$x_{\rm hard}$
model significantly over-predicts the observed slope for U+U. This
indicates that the variation in multiplicity between tip-tip
collisions and body-body collisions is smaller than anticipated if
multiplicity has a significant contribution proportional to $N_{\rm
  bin}$. Given this failure, we investigate two alternatives with no
explicit $N_{\rm bin}$ dependence: a constituent-quark Glauber model
(Glauber-CQ)~\cite{Eremin:2003qn,Adler:2013aqf} and the IP-Glasma
model~\cite{Schenke:2014tga} based on gluon
saturation~\cite{McLerran:1993ni}. The Glauber-CQ model neglects
$N_{\rm bin}$ and counts the number of participating constitutent
quarks $N_{\rm CQ}$ with each nucleon being treated as three
constituent quarks distributed according to $\rho=\rho_{0}\exp(-ar)$
with $a=4.27$ fm$-1$~\cite{Adler:2013aqf}. This model with
$\sigma_{qq}=9.36$ mb provides a good description of transverse energy
and multiplicity distributions at RHIC~\cite{Adler:2013aqf} and a
better description of $v_2$ fluctuations than a nucleon based Glauber
model~\cite{Agakishiev:2011eq}. In our simulation, for each $N_{\rm
  CQ}$, we sample an NBD with parameters tuned to match the
distributions from p+p~\cite{Ansorge:1988kn} and Au+Au at 200 GeV
($n=0.76$, and $k=0.34$ for $|\eta|<0.5$ and $n=2.9$ and $k=0.86$ for
$|\eta|<1$). For both Glauber models we use two sets of parameters for
the nuclear geometry, one corresponding to the more commonly used
values~\cite{Masui:2009qk} (dashed lines) and the new parameters
proposed in Ref.~\cite{NewUU} (solid lines). The effect of the
different parameter sets is small. The IP-Glasma and Glauber-CQ model
are also compared to the Au+Au data (Glauber-$x_{\rm hard}$ is left off
for clarity) but because of significant uncertainty in the actual
shape of a Au nucleus, it is difficult to draw conclusions from this
comparison.

In U+U collisions, both the IP-Glasma model and the Glauber-CQ model
predict slopes closer to the data. In the Glauber-CQ model, even
though there is no dependence on $N_{\rm bin}$, the average number of
quarks struck in a nucleon ($N_{\rm CQ}/N_{\rm part}$) is larger in
tip-tip than in body-body collisions so that tip-tip collisions create
more multiplicity. This leads to a strong anti-correlation between
$N_{\rm CQ}/N_{\rm part}$ and $\varepsilon_2$ which in turn translates
into a negative slope in $v_2$ vs. multiplicity. The IP-Glasma model
exhibits similar behavior. In gluon saturation models like the
IP-Glasma model, the multiplicity depends on
$Q_s^2S_{\perp}/\alpha_S(Q_s^2)$~\cite{Schenke:2014tga} where $Q_s^2$
(the saturation scale) is determined by the thickness of the nucleus
along the beam axis, $S_{\perp}$ is the transverse size of the overlap
region, and $\alpha_{S}$ is the strong coupling constant. For tip-tip
collisions, the increase in $Q_s^2$ in the numerator will be balanced
by a decrease of $S_{\perp}$. In the denominator, however, $\alpha_{S}$
decreases logarithmically with $Q_s^2$ leading to an increased
multiplicity in tip-tip collisions compared to body-body collisions.

The slope of $v_2$ vs. multiplicity provides a detailed probe of the
multiplicity production mechanism and the degree of coherence in
nuclear collisions. We find that accounting for the observed slope
seems to require models that include effects from sub-nucleonic
structure and significantly more coherence than is expected from the
Glauber-$x_{\rm hard}$ model. Previous studies questioned the
relevance of $N_{\rm bin}$ because of the apparent lack of an energy
dependence to $x_{\rm hard}$ and because the Glauber-CQ model also
provides a good description of multiplicity data. This study however,
provides direct evidence contradicting the Glauber-$x_{\rm hard}$
model.

In summary, we measured $v_2\{2\}$ and $v_2\{4\}$ for minimum bias,
and nearly fully overlapping Au+Au and U+U collisions at
$\sqrt{s_{NN}}=200$ and 193 GeV respectively. 
The knee structure in high multiplicity U+U collisions predicted by a
Glauber model with a two component multiplicity model with a
dependence on $N_{\rm bin}$ is not observed in $v_2$ versus $dN_{\rm
  ch}/d\eta$. Also, $v_2$ scaled by $\varepsilon_2$ from this model is
found to saturate and then decrease for the most central U+U
collisions. These findings indicate a weakness in the two-component
multiplicity calculation that is commonly used as part of Glauber
models in heavy ion collisions. We also used the STAR ZDCs to select
nearly fully overlapping collisions and showed that for a stringent
0.125\% ZDC selection criterion, the variation of $v_2$ with
multiplicity in U+U collisions is dominated by the different
geometries arising from the prolate shape of the uranium nucleus. This
demonstrates that ZDCs and multiplicity can be used to select tip-tip
or body-body enriched event samples. The variation of $v_2$ with
multiplicity in nearly fully overlapping collisions was shown to again
disfavor the Glauber model including a fractional contribution of
$N_{\rm bin}$ to multiplicity. Models with no explicit $N_{\rm bin}$
dependence such as a gluon saturation based model (IP-Glasma) or a
constituent quark Glauber model agree better with the data. In
addition to revealing fundamental information about the nature of
particle production in heavy-ion collisions, the findings in this
letter lay the groundwork for more extensive studies of the effect of
the initial geometry on other observables in nearly fully overlapping
collisions.

We thank the RHIC Operations Group and RCF at BNL, the NERSC Center at
LBNL, the KISTI Center in Korea, and the Open Science Grid consortium
for providing resources and support. This work was supported in part
by the Office of Nuclear Physics within the U.S. DOE Office of
Science, the U.S. NSF, the Ministry of Education and Science of the
Russian Federation, NNSFC, CAS, MoST and MoE of China, the Korean
Research Foundation, GA and MSMT of the Czech Republic, FIAS of
Germany, DAE, DST, and UGC of India, the National Science Centre of
Poland, National Research Foundation, the Ministry of Science,
Education and Sports of the Republic of Croatia, and RosAtom of
Russia.

\end{document}